# Neutron absorption correction and mean path length calculations for multiple samples with arbitrary shapes applications to highly absorbing samples on the Multi-Axis Crystal Spectrometer at NIST

Jose A. Rodriguez-Rivera[a,b]* and Chris Stock[c]

[a]*Department of Materials Science, University of Maryland, College Park, Maryland 20742 United States, [b]NIST Center for Neutron Research, National Institute of Standards and Technology, 100 Bureau Dr., Gaithersburg, Maryland 20899, United States, and [c]School of Physics and Astronomy, University of Edinburgh, Edinburgh EH9 3JZ, United Kingdom. E-mail: jose.rodriguez@nist.gov*

## Abstract

The finite volume algorithm for absorption correction developed by Wunch and Prewitt is examined. This algorithm is based on the numerical integration of the transmission function where three-dimensional quadratic surfaces define the sample boundaries. The algorithm can also be used to calculate the mean path length required for second-extinction calculations. We apply this method to the neutron inelastic scattering measurements of CeRhIn$_5$ using the Multi-Axis Crystal Spectrometer (MACS) at NIST. The algorithm has been expanded to correct the absorption of multiple coaligned samples. We show that this procedure can account for the angular-dependent absorption, and the technique can be used to correct data and plan experiments.

# 1. Introduction.

The developments of high-intensity neutron sources offer a dramatic improvement in the ability to probe the dynamic properties of materials at the nanoscale. The new high-intensity neutron spectrometers allow us to study smaller samples and samples containing atoms with large absorption cross-sections. Understanding the angular dependency of absorption corrections in neutron scattering experiments is increasingly important for comprehending the properties of materials. In particular, many strongly correlated electronic systems contain highly absorbing samples, which will give a strong angular dependence to the magnetic cross-section and are not inherent to the system properties of interest. Examples include materials containing elements such as Iridium ($Na_2IrO_3$), Indium ($CeCoIn_5$ and $CeRhIn_5$), Rhodium ($YbRh_2Si_2$), and others (Fig. 1). Many of these systems have been studied recently due to neutron instrumentation developments. While Fig. 1 illustrates that the absorption decreases with increasing energy, the physical properties of interest in many of these materials require fine energy resolution and hence low energies where the absorption cross section increases dramatically.

The absorption corrections for such materials are significant because the data are notably affected by a combination of neutron transmission (attenuation) and sample geometry, leading to misleading interpretations. Several techniques have been developed to calculate it quickly and accurately. The transmission factors calculated by Rouse et al. (Rouse *et al.*, 1970) are still widely used due to their easy accessibility in the crystallographic tables but are limited only to cylinders and sphere-shaped samples. Busing and Levy (Busing & Levy, 1957) developed one of the first numerical methods to calculate the absorption for an arbitrary shape. Meulenaer and Tompa (Meulenaer & H., 1965; Alcock *et al.*, 1972) method divides the sample into Howells polyhedra and applies an analytical formula to calculate the absorption. The advantage of such a method is the fast computing time to calculate the transmission for a single sample and detector position. Still, the sample must be divided into Howells polyhedral for every sample rotation and detector configuration. Wuensch and Prewitt (Wuensch & Prewitt, 1965) finite volume algorithm is a generalization of the Busing and Levy (Busing & Levy, 1957) technique that only requires the sample to be approximated by quadratic plane surfaces. Other absorption correction algorithms have been developed for specific cases (Angel, 2004; Blessing, 1995; Bowden & Ryan, 2010; Dallmann *et al.*, 2024; Hermann & Emrich, 1987; Lu *et al.*, 2024; Montesin *et al.*, 1991; Schmitt & Ouladdiaff, 1998). The project MANTID comprises a collection of absorption correction algorithms, including a calculator for an arbitrary sample defined as a collection of cuboids (Arnold *et al.*, 2014). This paper will review an updated version of the generalized Wunch and Prewitt algorithm to be applied to inelastic neutron scattering data. The algorithm has been extended to calculate the absorption for multiple coaligned samples. The same method can be used to find the neutron mean path length to calculate the Zachariasen extinction correction (Becker & Coppens, 1974).

## 2. The method.

The transmission is defined as $T = I/I_0$, where $I_0$ is the intensity of the incident beam and $I$ is the intensity of the diffracted beam from a crystal. The transmission coefficient $T$ is given by,

$$\langle T \rangle = \frac{1}{V} \int_V e^{-(\mu t + \mu_0 t_0)} dV. \tag{1}$$

Where $t$ and $t_0$ are the path lengths of the scattered and incident beam, $\mu$ and $\mu_0$ are the linear absorption coefficients for the incident and the scattered beam, respectively, and $dV$ is a given volume element. We replace eq. 1 for a discrete sum over all $(x_p, y_q, z_r)$ volume elements by its discrete equivalent:

$$\langle T \rangle = \frac{1}{M} \sum_{pqr} e^{-(\mu t + \mu_0 t_0)}. \qquad (2)$$

We can find the mean path length used for the Zachariasen extinction correction, which is defined as $\overline{T} = \frac{1}{T}\frac{\delta T}{\delta \mu}$ (Coppens & Hamilton, 1970), where,

$$\overline{T} = \frac{1}{M} \frac{\sum_{pqr}(t+t_0)e^{-\mu(t+t_0)}}{\sum_{pqr} e^{-\mu(t+t_0)}}. \qquad (3)$$

*2.1. Discretization of the volume.*

To evaluate numerically eqs. 2 and 3, we divide our sample into volume elements. The most common discretization method divides the volume into finite and regular rectangular cuboids. The method consists of a finite partitioning of the volume $V$. We define a box containing the sample as below:

$$x_{min} \leq x \leq x_{max},$$

$$y_{min} \leq y \leq y_{max},$$

$$z_{min} \leq z \leq z_{max}. \qquad (4)$$

Then, we subdivide our box into desired intervals $N_x$, $N_y$ and $N_z$. The array of grid points will be defined as a collection of indices p, q, and r. Each volume element center position will be given by,

$$\begin{aligned}
x_p &= x_{min} - \frac{x_{max} - x_{min}}{2N_x} + p\frac{x_{max} - x_{min}}{N_x}, \\
y_q &= y_{min} - \frac{y_{max} - y_{min}}{2N_y} + q\frac{y_{max} - y_{min}}{N_y}, \\
z_r &= z_{min} - \frac{z_{max} - z_{min}}{2N_z} + r\frac{z_{max} - z_{min}}{N_z}.
\end{aligned} \qquad (5)$$

*2.2. Finding t and $t_0$.*

The main problem in solving eq. 2 and eq. 3 is to find the neutron path $t$ and $t_0$ for each $\Delta V$ element. We can simplify our problem by defining our sample as a collection of one or more surfaces $f(x,y,z)$. Each surface has a quadratic form:

$$f(x,y,z) = Ax^2 + By^2 + Cz^2 + Dxy + Eyz + Gzx + Hx + Iy + Jz = F. \qquad (6)$$

The defined sample must not have re-entrant angles, meaning any volume element inside the sample must have a direct line of sight to the other volume elements. A point inside the sample must always be defined. This point will help determine whether any volume element is localized inside or outside the sample volume when the quadratic surfaces defining the sample contain two or more enclosed volumes, such as a sphere bisected by a plane. The transmission for samples with re-entrant angles can be calculated by constructing a collection of smaller samples with no re-entrant angles and applying the algorithm to calculate the transmission for multiple samples, as explained in the following sections.

To find the distance path that the neutron travels inside the sample $t$ and $t_0$, we define our scattered and incident neutron beam unit vectors $\mathbf{s}$ and $\mathbf{s_0}$ as in Fig. 2,

$$\mathbf{s} = s_x\hat{\mathbf{x}} + s_y\hat{\mathbf{y}} + s_z\hat{\mathbf{z}},$$

$$\mathbf{s_0} = s_{0x}\hat{\mathbf{x}} + s_{0y}\hat{\mathbf{y}} + s_{0z}\hat{\mathbf{z}}. \tag{7}$$

If a neutron travels a $t_0$ distance from a sample boundary $t_o$ a volume element $(x_p, y_q, z_r)$ and a $t$ distance from the same volume element to the respective sample boundary following the path defined by $\mathbf{s_0}$, and $\mathbf{s}$, we have:

$$x - x_p = -t_0 s_{0x}, \qquad x - x_p = t s_x,$$

$$y - y_q = -t_0 s_{0y}, \qquad y - y_q = t s_y,$$

$$z - z_r = -t_0 s_{0z}, \qquad z - z_q = t s_z. \tag{8}$$

Note that $t_0$ is defined as a negative multiple of $\mathbf{s_0}$ to make the positive $t_0$ the correct distance once the below quadratic function is solved. The substitution of eqs. 7 and 8 in to the bounding functions, eq. 6, leads to:

$$U_0 t_0^2 + V_0 t_0 + W_0 = 0, \tag{9}$$

with:

$$U_0 = As_{0x}^2 + Bs_{0y}^2 + Cs_{0z}^2 + Ds_{0x}s_{0y} + Es_{0y}s_{0z} + Gs_{0z}s_{0x}, \tag{10}$$

$$V_0 = -[2Ax_ps_{0x} + 2By_qs_{0y} + 2Cz_rs_{0z} + D(x_ps_{0y} + y_qs_{0z}) +$$

$$E(y_qs_{0z} + z_rs_{0y}) + G(z_rs_{0x} + x_ps_{0z}) +$$

$$Hs_{0x} + Is_{0y} + Js_{0z}], \tag{11}$$

$$W_0 = Ax_p^2 + By_q^2 + Cz_r^2 + Dx_py_q + Ey_qz_r + Gz_rx_p +$$

$$Hx_p + Iy_q + Jz_r - F. \tag{12}$$

Similarly, for $t$, we have

$$Ut^2 + Vt + W = 0, \tag{13}$$

where $U$ and $W$ are the same as in eq. 10 and 12 but the sign of $V$ is positive, opposite to $V_0$ as defined in eq. 11. Eqs. 9 and 13 have two solutions for $t_0$ and $t$, where the positive solution is always the correct one. In the case of the bounding function eq. 6 is a plane, then

$$t_0 = -W_0/V_0, \qquad t = -W/V. \tag{14}$$

The volume element must satisfy the next conditions to be evaluated in eqs. 2 and 3: 1) $t_0$ and $t$ must have positive values, and 2) The volume elements must also be connected to any other volume element inside the sample with a straight line without the interference of a boundary function. All other volume elements must be ignored. A collection of $n$ surfaces with the form of eq. 6 could result of more than one $t_0$ and $t$ positive values. In that case, we always take the smallest positive value of $t_0$ and $t$ that satisfy the second condition.

*2.3. Sample Orientation and kinematic equations*

Once the sample boundary functions are defined, we set the sample orientation to match the initial scattering conditions. To simplify our calculations, we will fix our sample and only rotate the incident and scattered beam unitary vectors **s₀** and **s**. We must note that the beam will rotate in the opposite direction of the sample. Depending on the diffractometer's geometry, the diffraction transformation can be applied (Busing & Levy, 1967; Thorkildsen *et al.*, 1999). For the examples in this paper, we will utilize rotation transformations about the *x*, *y*, and *z* axes as necessary.

For a typical triple-axis-spectrometer or a time-of-flight neutron spectrometer, where the measured scattering process takes place in the *x* – *y* plane, we define $\mathbf{Q}_x$ perpendicular to the incident beam and $\mathbf{Q}_y$ in the direction of the incident beam $\mathbf{s}_0$ and $\theta_{sample} = 0$. We can directly apply the kinematic equations to calculate the neutron transmission of the sample as a function of reciprocal space. For convenience, we will apply the transformation operations to the incident $\mathbf{s}_0$ and the scattered beam $\mathbf{s}$ rather than to the boundary functions. For a certain $A_3$ (defined as the sample rotation in the scattering plane $\theta_{sample}$) and $A_4$ (defined as the detector angle in the scattering plane $2\theta_{sample}$) sample position. we define kinematic equations as,

$$|\mathbf{Q}|^2 = \mathbf{k_i}^2 + \mathbf{k_f}^2 - 2|\mathbf{k_i}||\mathbf{k_f}|\cos A_4, \tag{15}$$

$$\omega = (\pm)\arccos\left(\frac{\mathbf{k_i}^2 + \mathbf{Q}^2 - \mathbf{k_f}^2}{2|\mathbf{k_i}||\mathbf{Q}|}\right) + (\pi/2 - A_3), \tag{16}$$

$$\mathbf{Q} = |\mathbf{Q}|(\cos\omega\ \hat{\mathbf{i}} + \sin\omega\ \hat{\mathbf{j}}). \tag{17}$$

Where $\mathbf{k_i}$ and $\mathbf{k_f}$ are the initial and final wavevectors, and $\mathbf{Q} = \mathbf{k_i} - \mathbf{k_f}$. The ± factor in the $\omega$ equation corresponds to ± $A_4$ values.

*2.4. Simple geometrical examples: spheres, cylinders, half cylinders.*

The program repository is available in MATLAB, and PYTHON. The repositories contain several pre-defined samples, including spheres, cylinders, half-cylinders, prisms, and cuboids. This paper will show examples of transmission calculations for a sphere and cylinder as a comparison with the IUCr international tables and the transmission calculation in the full range of $\theta$–$2\theta$ rotation for a half-cylinder sample. We define the sphere sample with one boundary function as in eq. 6 with $A = 1, B = 1, C = 1$ and $F = r^2$. All other values are zero. The linear absorption coefficient value is $\mu r = 2.5$ (Fig.3).

The cylinder sample has a radius $r = 1cm$ and a height of $l=0.5cm$. Three boundary functions define the cylinder as in eq. 6. The first boundary equation is a circle in the x-y plane with values of $A = 1$, $B = 1$, and $F = r^2$. All the other values are zero. Two planes dissect the cylinder and define the height. The first plane $z = l/4$ is defined with $J = 1$ and $F = l/4$. The other plane, $z = -l/4$, is defined by $J = 1$ and $F = -l/4$. The scattering plane lies in the x-y direction. The linear absorption correction is defined as $\mu r = 2.5$ to compare with the IUCr tables. For both samples, we used a grid of $N_x = N_y = N_z = 55$. The transmission calculated for the sphere and the cylinder completely agrees with the results in the IUCr tables.

The half-cylinder is an infinite cylinder with a radius of $1cm$ dissected by three planes: one at $y = 0$ and two at $z = 0.5$ and $z = -0.5$, respectively. The linear absorption coefficient $\mu = 2cm^{-1}$. Fig. 4 illustrates the relationship between the transmission, sample rotation angle $\theta$, and detector position $2\theta$. For this example, $\theta = 0°$ when the $y$ axis is parallel to the $s_0$ incident beam. $\theta$ and detector position $2\theta$ rotate in the same direction.

## 2.5. Multiple samples.

To calculate the total transmission for $N$–samples, we define each $n$ sample as a collection of quadratic surfaces as in eq. 6. Each sample must have a defined point inside to define the volume of interest. Each $n$ sample is divided into volume elements with the shape of regular rectangular cuboids as in eq. 5. For a volume element in the $n^{th}$ sample, the distances $t^n_{0,pqr}$ and $t^n_{pqr}$ inside the sample are calculated as described in section 2.2. We add the distances traveled inside the other samples in the paths of the incident $s_0$ and scattered $s$ beams (Fig. 5).

The beam paths traveled inside the samples are defined as $t^{m,n}_{0,pqr}$ and $t^{m,n}_{pqr}$, where $m$ is the sample number outside the corresponding sample $n$ ($m \neq n$) in the directions of the incident and scatted beam direction $s_0$ and $s$. To calculate the distances $t^{m,n}_{0,pqr}$ and $t^{m,n}_{pqr}$, First, we calculate the distances to the first boundary functions for the $m$ sample $t^{s0}_n$ and $t^s_n$ and add a small increment $\delta$. Then we define the points $(x^{s0}_n, y^{s0}_n z^{s0}_{ns})$ and $(x^s_n, y^s_n z^s_n)$ as,

$$x^{s0}_n = -(t^{s0}_n + \delta)s_{0x} + x_p, \quad x^s_n = (t^s_n + \delta)s_x + x_p,$$
$$y^{s0}_n = -(t^{s0}_n + \delta)s_{0y} + y_q, \quad y^s_n = (t^s_n + \delta)s_y + y_q,$$
$$z^{s0}_n = -(t^{s0}_n + \delta)s_{0z} + z_r, \quad z^s_n = (t^s_n + \delta)s_z + z_r. \tag{18}$$

.

We then test whether the points $(x^{s0}_n, y^{s0}_n z^{s0}_n)$ and $(x^s_n, y^s_n z^s_n)$ are located inside the $m^{th}$ samples. If the points are not located inside the sample $m^{th}$, we iterate them to the next boundary of the $m$ sample. If the volume elements are located inside the $m^{th}$ sample, then we calculate the path length $t^{m,n}_{0,pqr}$ in the $s_0$ path and $t^{m,n}_{pqr}$ in the $s$ path. We repeat this procedure for all the samples and the $(x_p, y_q, z_r)$ volume elements of all $N$ samples where $n \neq m$. We evaluate the contribution of all the samples in each volume element,

$$\langle T \rangle = \frac{1}{M} \sum_{n=1}^{N} \sum_{pqr}^{V_n} e^{-[\mu(t^n_{pqr} + \sum_{m \neq n} t^{m,n}_{pqr}) + \mu_0(t^n_{0,pqr} + \sum_{m \neq n} t^{m,n}_{0,pqr})]}, \tag{19}$$

where $M$ is the total number of volume elements inside all the $N$ samples. Fig. 6 shows the transmission algorithm of an array of three samples: a cube of $l$ = 2*cm* with the center located at (2,2,2), a sphere with $r$ = 1*cm* located at (−2,−2,−2) and a cylinder with $r$ = 1*cm* and a height of $l$ = 1.5*cm* located at (−2,2,0). the linear absorption correction $\mu$ = 2.0*cm*$^{-1}$. The incident beam points the direction of the y-axis, $s_0$ = $\hat{\mathbf{y}}$ when $\theta$ = 0. The sample rotation $\theta$ is defined as positive when it rotates clockwise, as the instrument CHRNS-MACS.

### 3. The case of CeRhIn$_5$.

CeRhIn$_5$ is an itinerant antiferromagnet with a tetragonal nuclear structure and a helical magnetic structure. The compound is structurally related to CeCoIn$_5$, which has the highest known heavy fermion superconducting transition temperature (T$_c$=2.3 K), and inelastic scattering measurements have found a strong coupling between magnetic and superconducting fluctuations (Stock *et al.*, 2015; Brener *et al.*, 2024). In contrast to CeCoIn$_5$, the Rh variant is not a superconductor, except possibly at very low temperatures, and displays spatially long-range antiferromagnetic order. To formulate a dispersion relation, the MACS spectrometer was used to map out the momentum dependence of the magnetic fluctuations at a series of energy transfers(Rodriguez *et al.*, 2008). The momentum dependence of the magnetic scattering intensity is linked to the direction of the moment fluctuations. This quantity is highly angular dependent, and therefore, it is central to understanding the sample absorption to, in turn, understand the underlying physics in this system.

Indium has a large absorption cross-section (193.8 barns at E=25 meV), and the linear attenuation factor decreases with the incident neutron beam energy from $\mu$=10.36 cm$^{-1}$ at E=2.4 meV up to $\mu$=4.04 cm$^{-1}$ at E=16 meV (Sears, 1992). Rhodium has a similar absorption cross section (144.8 barns at E=25 meV), making scattering experiments on CeRhIn$_5$ particularly difficult over other compounds like CeCoIn$_5$, for example. The sample was modeled as a semi-cylinder intersected by five planes for absorption correction. Absorption effects strongly affect the dispersion features with

Fig. 7 (*a−b*) illustrating the neutron transmission for negative and positive 2$\theta$ angles. We can observe that the neutron transmission is below 2% due to the more considerable

11path distance traveled by the neutrons inside the sample. However, most of the data at high $2\theta$ values satisfy the scattering reflection geometry of the sample, where the path distance traveled by the neutrons inside the sample is shorter, giving higher neutron transmission values. The data has a solid angular dependence, showing the dispersion features of the sample at large $2\theta$. Fig. 8 shows how the neutron signal weakens with the neutron transmission of the sample.

## 4. Program repository.

The program repositories and documentation are available in MATLAB and Python on the following links:

https://github.com/macsatncnr/abscorrmatlab

https://github.com/macsatncnr/abscorrpython

## 5. Final remarks.

New neutron sources and instruments with high neutron flux have made it possible to study highly absorbent samples. A more careful data analysis must be performed when using this kind of sample due to the features that arise from neutron transmission combined with the sample shape. This algorithm can be easily implemented compared to others. It can be extended to complex cases like concave surfaces, capillaries, and samples that use different bulk materials by only re-interpreting the path-length solutions.

Access to CHRNS-MACS was provided by the Center for High-Resolution Neutron Scattering, a partnership between the National Institute of Standards and Technology and the National Science Foundation under Agreement No. DMR-1508249. Certain commercial software is identified in this paper to foster understanding. Such path distance traveled by the neutrons inside the sample. However, most of the data at high $2\theta$ values satisfy the scattering reflection geometry of the sample, where the path distance traveled by the neutrons inside the sample is shorter, giving higher neutron transmission values. The data has a solid angular dependence, showing the dispersion features of the sample at large $2\theta$. Fig. 8 shows how the neutron signal weakens with the neutron transmission of the sample.

## 4. Program repository.

The program repositories and documentation are available in MATLAB and Python on the following links:

https://github.com/macsatncnr/abscorrmatlab

https://github.com/macsatncnr/abscorrpython

## 5. Final remarks.

New neutron sources and instruments with high neutron flux have made it possible to study highly absorbent samples. A more careful data analysis must be performed when using this kind of sample due to the features that arise from neutron transmission combined with the sample shape. This algorithm can be easily implemented compared to others. It can be extended to complex cases like concave surfaces, capillaries, and samples that use different bulk materials by only re-interpreting the path-length solutions.

identification does not imply recommendation or endorsement by the National Institute of Standards and Technology, nor does it imply that this software is necessarily the best available for the purpose.

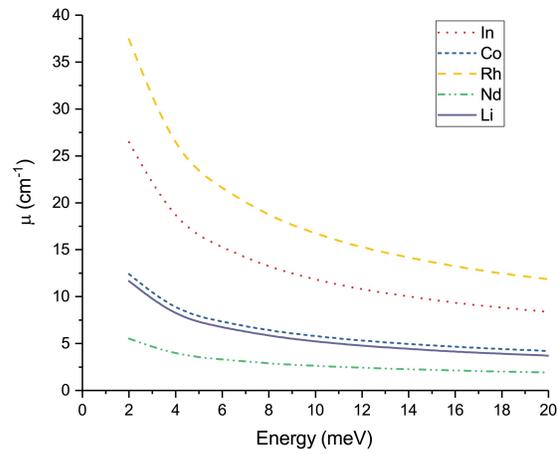

Fig. 1. Neutron linear attenuation factors for different elements as a function of energy (Sears, 1992).

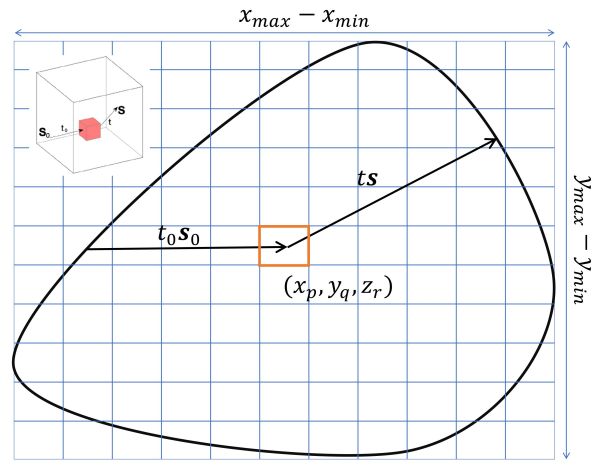

Fig. 2. Discretization of volume elements. The sample is divided into $N_x$, $N_y$, and $N_z$. Each square represents a volume element. $s_0$ and $s$ represent the incident and scattered beam unit vectors. $t_0$ and $t$ represent the incident and scattered beam path length inside the sample.

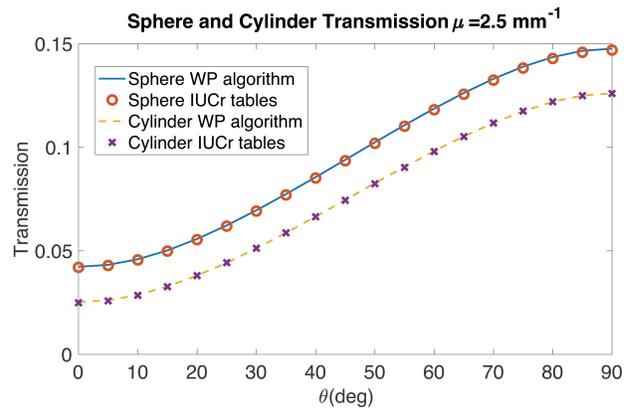

Fig. 3. Comparison of the calculated transmission for $\mu r$ = 2.5 of a sphere and a cylinder sample with the IUCr tables.

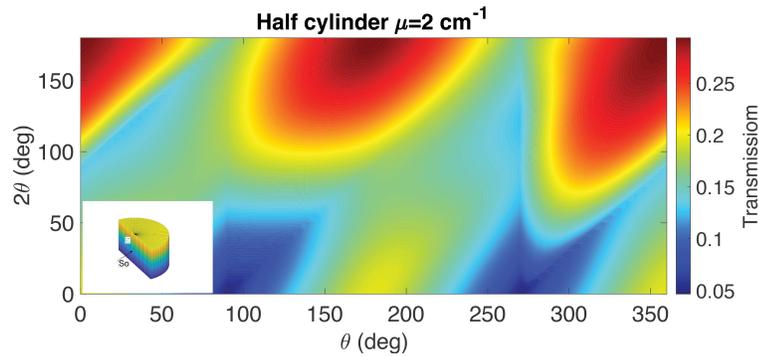

.

Fig. 4. Neutron transmission for a half cylinder $r = 0.5 cm$ as a function of the sample rotation $\theta$ and detector position $2\theta$. The arrow points to the $s_0$ incident beam when $\theta$ = 0º. The sample has a linear absorption coefficient factor of $\mu$ = 2.0 $cm^{-1}$.

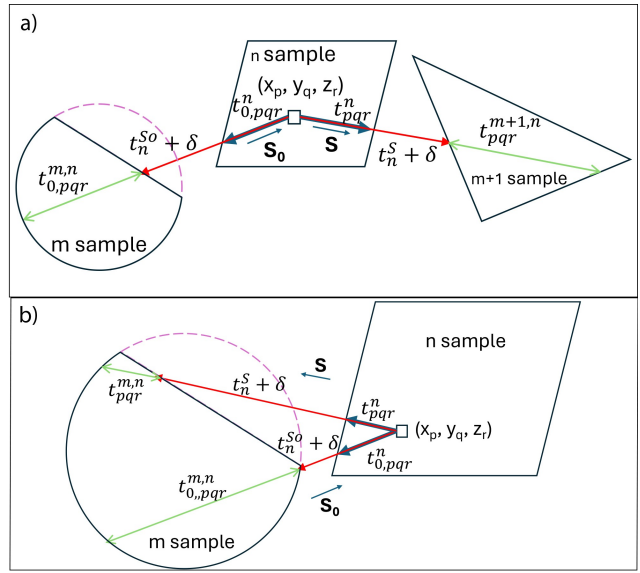

Fig. 5. a) Multiple samples configuration. In the direction of the $s_0$ incident beam, we define $t^n_{0,pqr}$ as the path distance from the $(x_p,y_q,z_r)$ volume element to the boundary function defining the $n^{th}$ sample, $t^{s0}_n + \delta$ as the distance from the volume element to the outside sample $m$ and $t^{m,n}_{0,pqr}$ the total path distance traveled inside the sample $m$. In the same way, we define for the $m + 1$ sample the distances $t^n_{pqr}$, $t^s_n + \delta$ and $t^{m+1,n}_{pqr}$ in the direction of the $s$ scattered beam ($m,m+1 /= n$). b) For a volume element $(x_p,y_q,z_r)$ inside the $n^{th}$ sample, the $m^{th}$ sample can be in located the way of the incident $s_0$ and scattered $s$ beam.

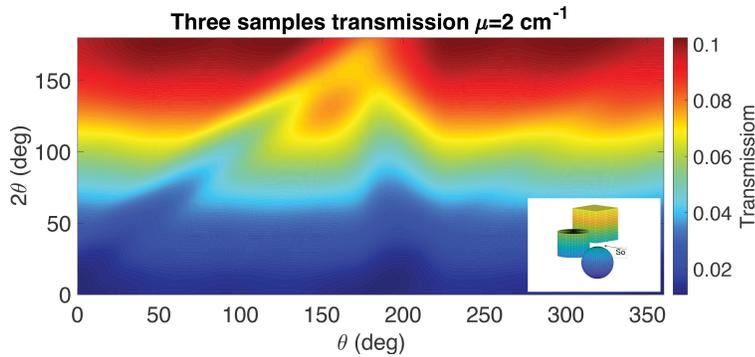

Fig. 6. Neutron transmission for a set of three samples: a cube, a sphere, and a cylinder. The rotation of $\theta$ and detector position $2\theta$ is defined as left-handed, which means that positive angles are clockwise, as with most NCNR instruments, like the CHRNS-MACS spectrometer.

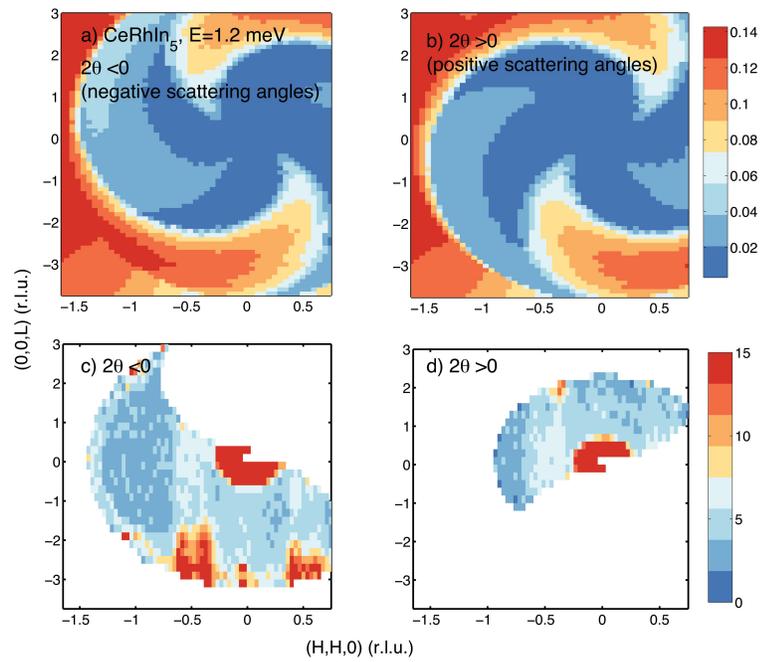

Fig. 7. CeRhIn$_5$ transmission at E=1.2meV (Stock *et al.*, 2015; Brener *et al.*, 2024). The total attenuation factor of the sample changes from $\mu$=7.33 cm$^{-1}$ at E=5.05 meV to $\mu$=5.17 cm$^{-1}$ at 9.65 meV (Sears, 1992). In this experiment, some of the relevant data is attenuated at small scattering angles.

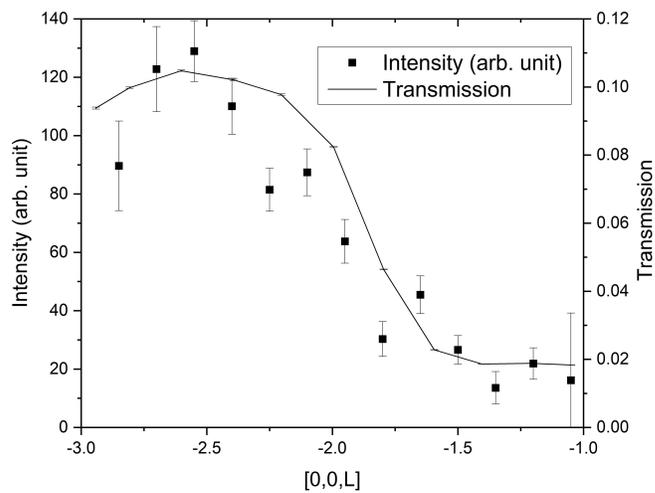

Fig. 8. CeRhIn5 [00L] neutron line scan at 0.6meV and neutron transmission. The feature shows a 1D spin interaction in the L direction. The intensity decay is mainly due to the neutron beam attenuation. Error bars represent one standard deviation.